# Strong Anomalous Optical Dispersion of Graphene: Complex Refractive Index Measured by Picometrology

*Xuefeng Wang, Yong P. Chen and David D. Nolte\**

Department of Physics, Purdue University, West Lafayette, IN 47907, USA

*Corresponding author: nolte@physics.purdue.edu

ABSTRACT

We apply spinning-disc picometrology to measure the complex refractive index of graphene on thermal oxide on silicon. The refractive index varies from n = 2.4-1.0i at 532 nm to n = 3.0-1.4i at 633 nm at room temperature. The dispersion is five times stronger than bulk graphite (2.67-1.34i to 2.73-1.42i from 532 nm to 633 nm).

Graphene is a two-dimensional atomic carbon crystal of a single layer of graphite, which can be prepared by micromechanical cleavage [1,2]. The unique structure of graphene gives it many distinguishing features, including field effect [2], anomalous quantum hall effect [3,4], high intrinsic strength [5] among others [6,7]. In contrast to the extensive electrical studies, optical studies of graphene have lagged because traditional optical techniques such as ellipsometry fail when applied to graphene because of graphene's sub-nanometer thickness, its dielectric anisotropy, small transverse sample size (usually 10 μm) and sparse distribution on the substrate when using the current manual exfoliation method. The optical dispersion of the refractive index, $\tilde{n}_g$, is a fundamental optical property of graphene that remains unsettled. Reflectometry has been applied to acquire reflectance spectra over a broad wavelength region to find an average $\tilde{n}_g$ by fitting the spectrum [8,9]. However, the results do not include dispersion. In this paper, we provide measurements of the dispersion of the refractive index at 488 nm, 532 nm and 633 nm.

Our experimental approach is spinning-disc picometrology, based on common-path interferometry [10], which can measure the complex refractive index of a thin film (down to 10 pm thickness) with small size (currently down to 4 μm) deposited on an flat substrate, using normal incidence at multiple wavelengths. Picometrology studies thin films deposited on a substrate that has a reflection coefficient $\tilde{r}$. The thin film (graphene in this paper) modifies $\tilde{r}$ of the substrate in both amplitude and phase. Traditional reflectometry only measures the amplitude change of $|\tilde{r}|$. Therefore it cannot acquire the complex change, which is essential for the calculation of complex $\tilde{n}$ of the thin film. With picometrology we measure the phase change by monitoring the asymmetric diffraction of the reflected beam when the focal spot scans over the edge of the thin film (Fig. 1). The center of the diffraction pattern in the far-field, or on a Fourier plane, is shifted when half the spot is on the substrate and the other half is on the graphene-modified substrate. Picometrology measures both amplitude and phase change of $\tilde{r}$ due to the thin film with a quadrant detector that simultaneously monitors both intensity and shift of a reflected beam.

The equations for picometrology treat a thin film on an arbitrary substrate. If the substrate has a reflection coefficient $\tilde{r}$ in air under conditions of wavelength $\lambda$ (in vacuum) and normal incidence, then after applying a thin film with thickness $d$ and refractive index $\tilde{n}_g$ on the substrate, the reflection coefficient is changed to $\tilde{r}'$ [10,11]

$$\tilde{r}' = \frac{(e^{i\delta_g} - e^{-i\delta_g})\tilde{r}_g + \tilde{r}(e^{-i\delta_g} - \tilde{r}_g^2 e^{i\delta_g})}{(e^{i\delta_g} - \tilde{r}_g^2 e^{-i\delta_g}) + \tilde{r}(e^{-i\delta_g} - e^{i\delta_g})\tilde{r}_g} e^{2i\delta_g/\tilde{n}_g} \qquad (1)$$

where $\delta_g = \frac{2\pi \tilde{n}_g}{\lambda} d$ and $\tilde{r}_g$ is the reflection coefficient at the interface between the ambient medium and the thin film.

If $d \ll \lambda$, Eq.1 is simplified to be

$$\tilde{r}' = \tilde{r} + \left[ \frac{(\tilde{r}_g - \tilde{r})(1 - \tilde{r}\tilde{r}_g)}{(1 - \tilde{r}_g^2)} + \frac{\tilde{r}}{\tilde{n}_g} \right] \frac{4\pi i \tilde{n}_g d}{\lambda} \qquad (2)$$

and using $\tilde{r}_g = \frac{1 - \tilde{n}_g}{1 + \tilde{n}_g}$, Eq. 2 is further simplified to be

$$\tilde{r}' = \tilde{r} + (1 + \tilde{r})^2 (1 - \tilde{n}_g^2) \frac{\pi i}{\lambda} d \qquad (3)$$

In summary, graphene with a thickness $d$ causes the reflection coefficient change $(1+\tilde{r})^2(1-\tilde{n}_g^2)\frac{\pi i}{\lambda}d$ on a substrate with original reflection coefficient $\tilde{r}$. The complex $\tilde{n}_g$ can be calculated by measuring both amplitude and phase change of $\tilde{r}$ caused by the graphene.

In Picometrology, a split detector is used to detect intensity (I) and phase-contrast (PC) signals which respectively carry the information of amplitude change and phase change of $\tilde{r}$ due to the graphene. The detector window is located on the Fourier plane of the reflected beam (Fig. 2). The detection window consists of two halves A and B with output signals I = B+A and PC=B-A.

The focal spot scans over the graphene sample with thickness d and profile $s(x)$ which is a Heaviside step function for the description of the graphene profile. The signals are normalized to $i^I(x)$ and $i^{PC}(x)$ by dividing the intensity change and the PC signal by the reflectance from the blank substrate. The relation between the normalized two-channel responses and the profile of the thin film (graphene) is

$$i^I(x) = \text{Im}\left[(\tilde{n}_g^2-1)\frac{(1+\tilde{r})^2}{\tilde{r}}\right]\frac{2\pi d}{\lambda}g^2(x)\otimes s(x)$$
$$i^{PC}(x) = \text{Re}\left[(\tilde{n}_g^2-1)\frac{(1+\tilde{r})^2}{\tilde{r}}\right]\frac{2\pi d}{\lambda}(D(x)\times g(x))\otimes s(x) \quad (4)$$

where $D(x)$ is Dawson function (Hilbert transformation of $g(x)$). The details of the derivations can be found in Ref. [10] [10]. The function $g(x)$ is a normalized Gaussian with $\int_{-\infty}^{\infty}g^2(x)dx=1$.

We compare the amplitudes of the I signal and the PC signal to find $\tilde{n}_g$. Eq. 4 can be re-expressed as

$$A[i^I(x)] = \text{Im}\left[(\tilde{n}_g^2-1)\frac{(1+\tilde{r})^2}{\tilde{r}}\right]\frac{2\pi d}{\lambda}A[g^2(x)\otimes s(x)]$$
$$A[i^{PC}(x)] = \text{Re}\left[(\tilde{n}_g^2-1)\frac{(1+\tilde{r})^2}{\tilde{r}}\right]\frac{2\pi d}{\lambda}A[(D(x)\times g(x))\otimes s(x)] \quad (5)$$

where A(f(x)) is the peak-to-peak value of f(x). The function $g^2(x)\otimes s(x)$ has amplitude unity wheras $(D(x)\times g(x))\otimes s(x)$ is a "pulse" function with an amplitude of 0.2805. Both amplitudes are independent of the width of $g(x)$, i.e. the focal size of the laser beam, giving

$$A\left[g^2(x) \otimes s(x)\right] = 1$$
$$A\left[(D(x) \times g(x)) \otimes s(x)\right] = 0.2805$$

This gives the equation for the refractive index measurement using picometrology to be

$$(\tilde{n}_g^2 - 1)\frac{(1+\tilde{r})^2}{\tilde{r}}\frac{2\pi d}{\lambda} = 3.565 A\left[i^{PC}(x)\right] + A\left[i^I(x)\right]i \quad (6)$$

For the data acquisition of graphene sheets on a silicon wafer, 2D scanning is performed on the graphene using a four-quadrant detector. The full PC response is acquired by quadrature summation of the two PC channels (Details are shown in supporting material). The coefficients of Eq. 6 remain the same and therefore the calibration for the Gaussian spot size is not required.

The picometrology scanning system is shown in Fig.2. The laser wavelength is 532 nm or 633 nm. The laser beam passes through a polarizer and a λ/4 waveplate for laser isolation, and passes through a beam expander (3×) to acquire high resolution. The expanded laser beam is focused on the sample by a 5× objective lens with a resolution of 2 μm. The silicon wafer is mounted on a spinner fixed on a linear stage. The 2D scans are performed by spinning the sample and translating the stage (resolution is 0.1 μm). The reflected light from the sample is collected by the lens and transformed to a quadrant detector at the Fourier plane. Two lenses and a pinhole are placed before the detector as a spatial filter. The quadrant detector has three outputs: the total intensity and two phase contrast signals.

Two graphene monolayer samples were tested at 532 nm and 633 nm wavelengths. (Sample 1 is mechanically exfoliated from natural graphite, and Sample 2 is purchased from Graphene Industries Company where graphene is made from highly oriented pyrolytic graphite (HOPG). Monolayers were previously identified on these samples). The substrates were silicon wafers grown with $SiO_2$ film. The $SiO_2$ thicknesses were measured to be 310 nm for sample 1 and 285 nm for sample 2 by analyzing the reflectance spectrum. The refractive index of $SiO_2$ and silicon is 1.457 and 3.78 at 532 nm, and 1.463 and 4.17 at 633 nm, respectively [12, 13]. The complex-valued reflection coefficients of samples 1 and 2 on their surfaces were calculated to be $\tilde{r}$ = -0.075 - 0.460i and $\tilde{r}$ = 0.270 - 0.211i at 532 nm and $\tilde{r}$ = 0.223 + 0.237i and $\tilde{r}$ = -0.047 + 0.429i at 633 nm. Fig.3 a1 and b1 show the two graphene samples observed under a microscope. In Fig.3 a2, a3, b2 and b3 are the 2D images of the I and PC channel data. The PC channel detects the graphene edge whereas the I channel detects the flat graphene regions.

In Fig.3 a4, the amplitudes of the normalized I and PC signals from the graphene monolayer are A(I) = (180.1-180.95)/180.95 = -0.0047 and A(PC) = 2.70/180.95 = 0.0149 (Fig. 3). Using the substrate reflectance, the thickness 0.335 nm of the graphene monolayer [8, 14, 15], 532 nm wavelength, and the measured amplitudes, the graphene refractive index is found to be $ñ_g$ = 2.37-0.97i for sample 1 and $ñ_g$ = 2.37-1.07i for sample 2 at 532 nm. We also acquired the complex refractive indexes of both samples at 633 nm wavelength (Fig. 4). The results give $ñ_g$ = 2.95-1.32i for sample 1 and $ñ_g$ = 2.98-1.44i for sample 2 at 633 nm. These results are listed in Table 1. The error of $ñ_g$ detection is estimated to be ±0.1 for both the real part and the imaginary part of $ñ_g$. The error sources are the imperfect Gaussian profile of the focused spot, slight defocus and local variations of the $SiO_2$ thickness.

For comparison, the refractive index of bulk graphite (many stacked layers of graphene) is 2.67-1.34i at 532 nm (2.33 eV) and 2.73-1.42i at 633 nm (1.96 eV) polarized perpendicular to the c-axis [12, 16]. With the averaged values for graphene refractive indexes $ñ_g$ = 2.4 – 1.0i and $ñ_g$ = 3.0-1.4i, respectively, the dispersion is five times greater for both the real part and imaginary parts. This strong dispersion of graphene is first reported and it is likely caused by the strongly modified quantum level structure of one atom layer compared to graphite. We also measured ñ of graphene bilayers and ñ of both monolayer and bilayer at 488 nm. The results are shown in Fig. 5 along with ñ of the bulk graphite [12].

In summary, picometrology provides a valuable and easily generalizable tool for the study of the optical properties of graphene. It measures the complex refractive index of graphene as a function of wavelength. It can also be applied to other research, such as micro-scale samples with similar limitations to the graphene (e.g., nanotube 2D array [17, 18]). Considering that many nanotechnologies involve micron-scale samples on a well-defined substrate, picometrology has the potential to provide a standard optical approach for the study of nanolayered structures.

**Acknowledgement.** This work was sponsored under grants from Quadraspec, Inc. and from the Indiana Economic Development Corporation through the Purdue Research Foundation. We also thank Dr. Liyuan Zhang for preparation of sample 1 and Prof. Jiangping Hu for his insightful interpretation of the electronic properties of graphene.

**References:**

1. Novoselov, K. S.; Jiang, D.; Schedin, F.; Booth, T. J.; Khotkevich, V. V.; Morozov, S. V.; Geim, A. K. *Proceedings of the National Academy of Sciences of the United States of America* **2005,** 102, (30), 10451-10453.

2. Novoselov, K. S.; Geim, A. K.; Morozov, S. V.; Jiang, D.; Zhang, Y.; Dubonos, S. V.; Grigorieva, I. V.; Firsov, A. A. *Science* **2004,** 306, (5696), 666-669.

3. Novoselov, K. S.; Geim, A. K.; Morozov, S. V.; Jiang, D.; Katsnelson, M. I.; Grigorieva, I. V.; Dubonos, S. V.; Firsov, A. A. *Nature* **2005,** 438, (7065), 197-200.

4. Zhang, Y. B.; Tan, Y. W.; Stormer, H. L.; Kim, P. *Nature* **2005,** 438, (7065), 201-204.

5. Lee, C.; Wei, X. D.; Kysar, J. W.; Hone, J. *Science* **2008,** 321, (5887), 385-388.

6. Calogeracos, A. *Nature Physics* **2006,** 2, (9), 579-580.

7. Wang, F.; Zhang, Y. B.; Tian, C. S.; Girit, C.; Zettl, A.; Crommie, M.; Shen, Y. R. *Science* **2008,** 320, (5873), 206-209.

8. Ni, Z. H.; Wang, H. M.; Kasim, J.; Fan, H. M.; Yu, T.; Wu, Y. H.; Feng, Y. P.; Shen, Z. X. *Nano Letters* **2007,** 7, (9), 2758-2763.

9. Jung, I.; Pelton, M.; Piner, R.; Dikin, D. A.; Stankovich, S.; Watcharotone, S.; Hausner, M.; Ruoff, R. S. *Nano Letters* **2007,** 7, (12), 3569-3575.

10. Wang, X. F.; Zhao, M.; Nolte, D. D. *Applied Optics* **2007,** 46, (32), 7836-7849.

11. Heavens, O. S., *Optical Properties of thin solid films*. 1955; p 66~80.

12. *Handbook of optical constants of solids*. Academic Press: San Diego, 1991.

13. *Properties of Silicon*. INSPEC: New York, 1988.

14. Kelly, B. T., *Physics of Graphite*. Applied Science: London, 1981.

15. Dresselhaus, M. S.; Dresselhaus, G.; Eklund, P. C., *Science of Fullerenes and Carbon Nanotubes*. Academic Press: San Diego, 1996.

16. Djurisic, A. B.; Li, E. H. *Journal of Applied Physics* **1999,** 85, (10), 7404-7410.

17. Li, J.; Ng, H. T.; Cassell, A.; Fan, W.; Chen, H.; Ye, Q.; Koehne, J.; Han, J.; Meyyappan, M. *Nano Letters* **2003,** 3, (5), 597-602.

18. Hoyer, P. *Langmuir* **1996,** 12, (6), 1411-1413.

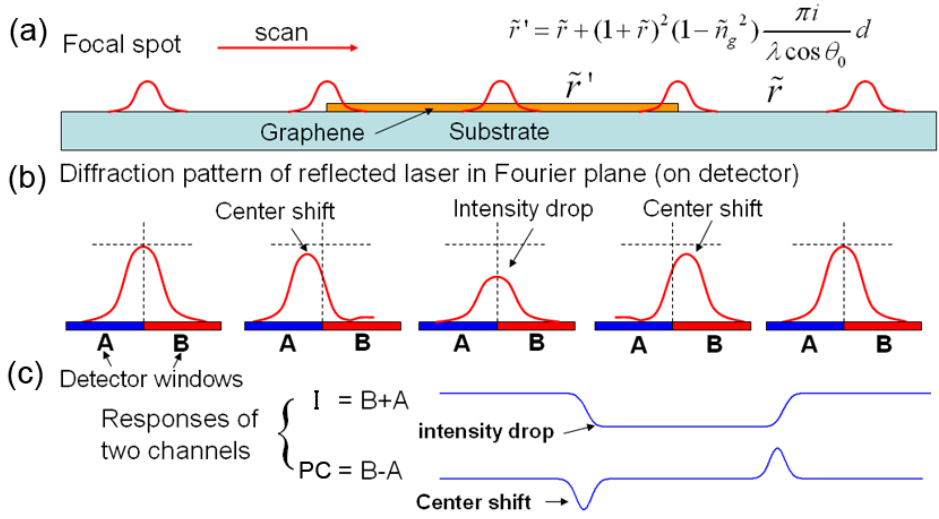

Figure 1. Principles of picometrology. A graphene layer modifies the reflection coefficient $\tilde{r}$ of a substrate into $\tilde{r}'$ according to Eq. 3. By measuring both the amplitude change and phase change of $\tilde{r}$, the complex ñg can be calculated. (a) A focused Gaussian beam scans across a graphene sample. The reflected light forms a Fraunhoffer diffraction pattern in the far field or Fourier plane. (b) When the spot scans across the edge of the graphene, the center of the diffraction pattern is shifted from the original position due to the phase difference between $\tilde{r}$ and $\tilde{r}'$. When the spot is on the graphene, the reflected intensity drops due to the amplitude difference between $\tilde{r}$ and $\tilde{r}'$. By combining the center shift and intensity drop of the diffraction pattern, the full change of $\tilde{r}$ is calculated. (c) A split detector is used to monitor the intensity drop and center shift simultaneously. The output I and PC signals are directly related with ñg by Eq. 6.

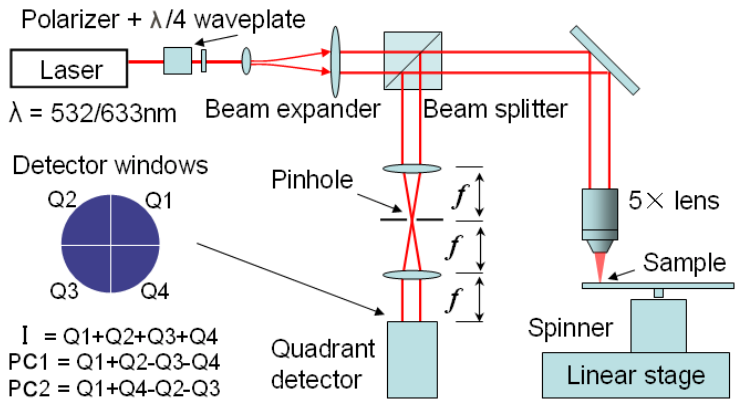

Figure 2. The optical layout of the Picometrology system.

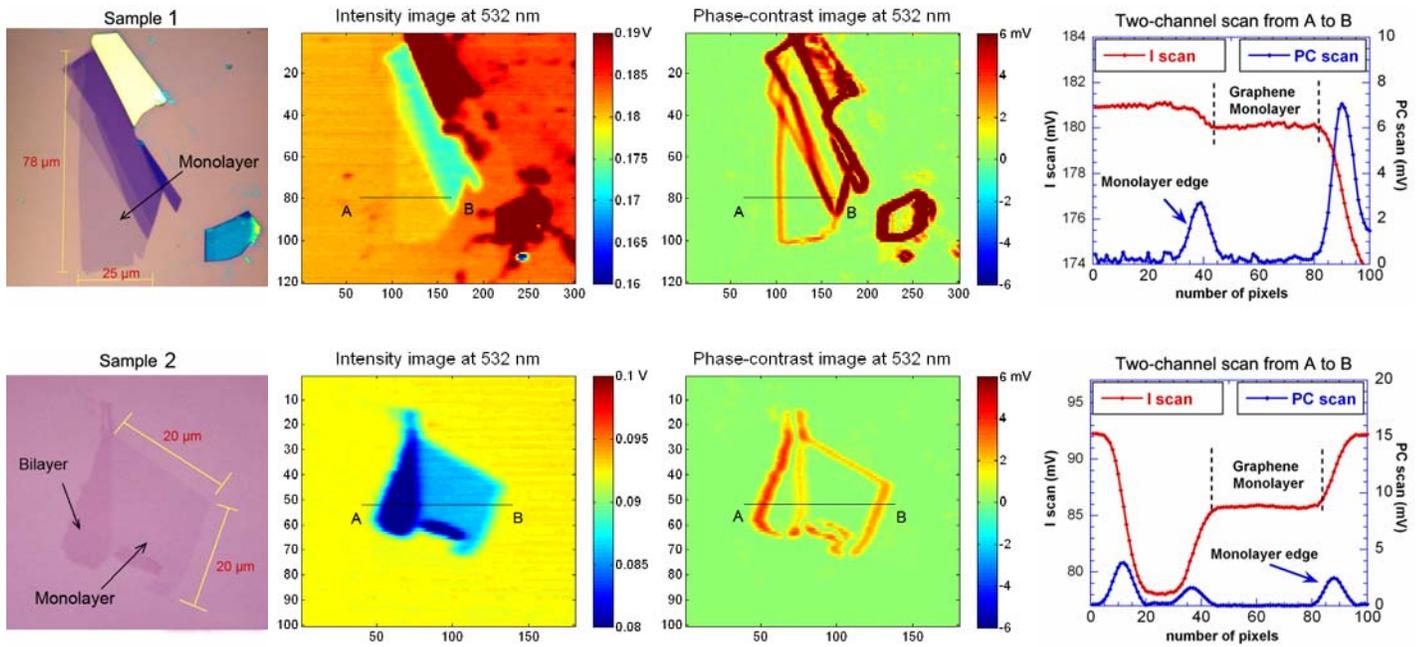

Figure 3. Two graphene samples scanned under 532 nm wavelength. The substrates for two graphene monolayer samples are silicon wafers grown with 310 nm and 285 nm SiO2 respectively. The normalized amplitudes of I and PC signals are calculated and listed in Table 1. Using Eq.6, ñ$_g$ was calculated to be 2.37-0.97i and 2.37 -1.07i for sample 1 and 2.

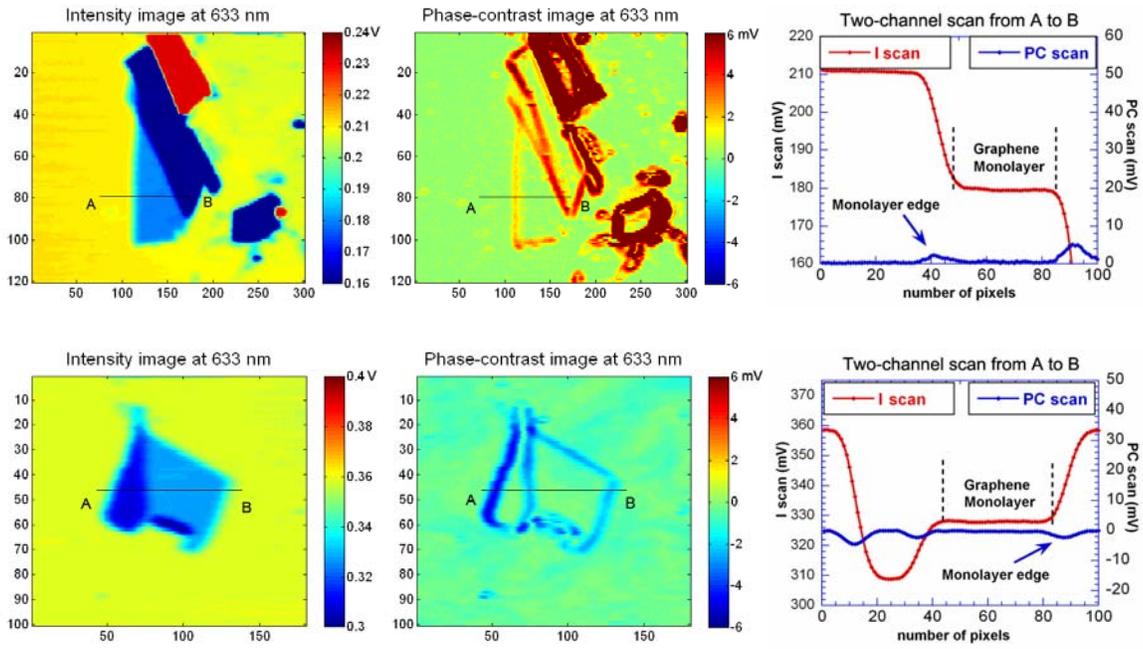

Figure 4. Two graphene samples scanned under 633 nm wavelength. The normalized amplitudes of I and PC signals are calculated and listed in Table 1. Using Eq. 6, $ñ_g$ was calculated to be 2.95-1.32i and 2.98 -1.44i for samples 1 and 2.

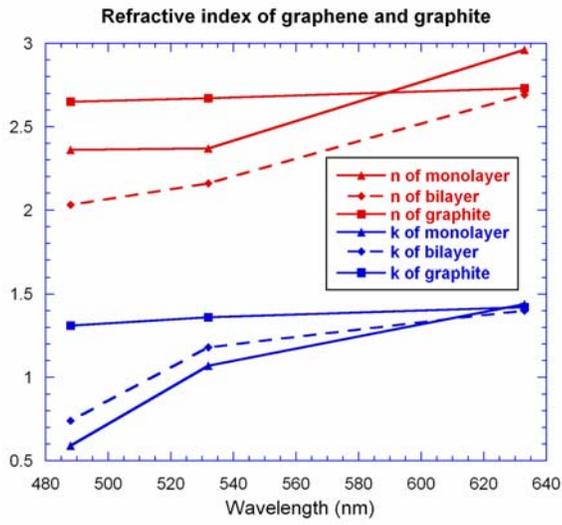

Figure 5. Complex refractive indexes of graphene monolayer, bilayer (based on the results of sample 2) and bulk graphite (ordinary refractive index).

Table 1. Measurements of complex refractive index of graphene.

|  | 488 nm | | 532 nm | | 633 nm | |
| --- | --- | --- | --- | --- | --- | --- |
|  | Sample 1 | Sample 2 | Sample 1 | Sample 2 | Sample 1 | Sample 2 |
| r of substrate | -0.499-0.328i | -0.111- 0.483i | -0.075-0.460i | 0.270-0.211i | 0.223+0.237i | -0.047+0.429i |
| $A(i^I)$ | 0.0132 | 0.0095 | -0.0047 | -0.07 | -0.1509 | -0.0851 |
| $A(i^{PC})$ | 0.0028 | 0.0124 | 0.0149 | 0.0267 | 0.0095 | -0.0059 |
| Measured $ñ_g$ | 2.64-0.56i | 2.36-0.59i | 2.37-0.97i | 2.37-1.07i | 2.95 – 1.32i | 2.98 -1.44i |

**Supporting Information**

Calculation method for the full PC signal:

The summed PC channel is acquired by PC = $(PC1^2 + PC2^2)^{1/2}$. The sign for the sum is based on image PC1. In our configuration of optical layout, the sign is positive if the PC1 response is negative when the Gaussian spot scans from the substrate onto the graphene, or positive from the graphene onto the substrate. The sign is negative if otherwise. The examples are provided as the following images. The sign of PC response is positive for sample 1 at 633 nm (Fig. a1, a2 and a3) while the sign is negative for sample 2 at 532 nm (Fig. b1, b2 and b3).

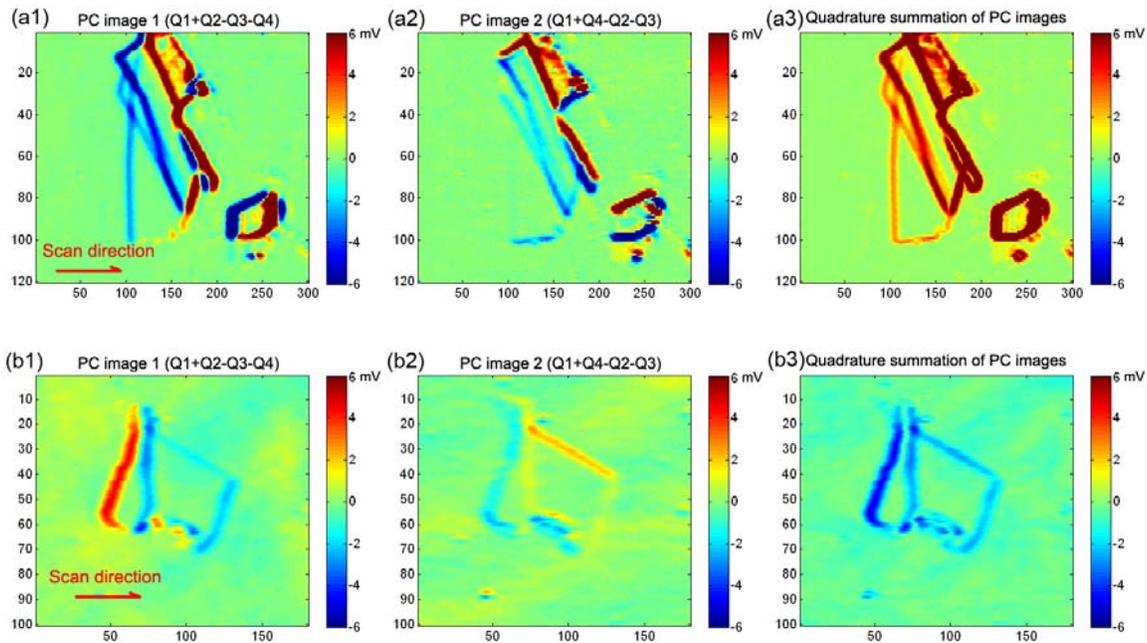